\documentclass[aps,superscriptaddress,amsmath,amssymb,floatfix,twocolumn,showpacs,amsfonts,longbibliography,prb]{revtex4-2}
\usepackage[english]{babel}
\usepackage{times}
\usepackage[varg]{txfonts}
\usepackage{textcomp}
\usepackage{graphicx}
\usepackage{subfigure}
\usepackage{tabu}
\usepackage{color}
\usepackage[plainpages=false,pdfpagelabels,colorlinks=true,linkcolor=red,urlcolor=blue,citecolor=blue,pdftitle={Title},pdfauthor={},pdfdisplaydoctitle=true,pdfduplex=DuplexFlipLongEdge]{hyperref}
\usepackage{braket}
\usepackage{float}
\usepackage{overpic}
\usepackage{gensymb}
\usepackage{mathrsfs}
\usepackage{tikz}
\usepackage{bm}
\usepackage{multirow}
\usepackage{amsfonts}
\usepackage{amsmath}
\usepackage{mathrsfs}
\usepackage{amssymb}
\usepackage{paralist}
\usepackage{indentfirst}
\usepackage{ulem}
\usepackage{soul}
\usepackage{cancel}
\usepackage{CJK}

\newcommand{\ketbra}[2]{\mathinner{|{#1}\rangle \langle{#2}|}}
\allowdisplaybreaks

\hypersetup{
           breaklinks=true,   
           colorlinks=true,   
           pdfusetitle=true,  
        }
        
\newcommand{\tcr}{\textcolor{red}}
\newcommand{\tcb}{\textcolor{blue}}

\begin{document}
\selectlanguage{english}
\title{Competition between dimerization and vector chirality in the spin-$3/2$ $J_1$-$J_2$ Heisenberg chain\\ with uniaxial single-ion anisotropy}

\author{Ji-Lu He}
\affiliation{Beijing Computational Science Research Center, Beijing 100084, China}

\author{Sebastian Eggert}
\email{eggert@physik.uni-kl.de}
\affiliation{Physics Department and Research Center OPTIMAS, RPTU Kaiserslautern-Landau, 67663 Kaiserslautern, Germany}

\author{Haiqing Lin}
\email[]{haiqing0@csrc.ac.cn}
\affiliation{Beijing Computational Science Research Center, Beijing 100084, China}
\affiliation{School of Physics and Institute for Advance Study in Physics, Zhejiang University, Hangzhou 310058, Zhejiang, China}

\author{Xiaoqun Wang}
\email[]{xiaoqunwang@zju.edu.cn}
\affiliation{School of Physics and Institute for Advance Study in Physics, Zhejiang University, Hangzhou 310058, Zhejiang, China}
\affiliation{Collaborative Innovation Center of Advanced Microstructures, Nanjing University, Nanjing 210093, China}

\author{Shi-Jie Hu}
\email{shijiehu@csrc.ac.cn}
\affiliation{Beijing Computational Science Research Center, Beijing 100084, China}
\affiliation{Department of Physics, Beijing Normal University, Beijing, 100875, China}

\begin{abstract}
The spin-$3/2$ chain is a versatile prototypical platform for the study of competition between different kinds of magnetic orders, with the objective of obtaining a deeper understanding of the corresponding quantum phase transitions.
In this work, we investigate the spin-$3/2$ chain with nearest-neighbor $J_1$, next-nearest-neighbor $J_2$, and uniaxial single-ion anisotropy $D$ terms in the absence of a magnetic field.
For positive values of $J_2/J_1$ and $D/J_1$, we find  seven different phases in a rich phase diagram.
Without frustration $J_2=0$, a gapless Luttinger liquid phase remains stable for all $D>0$.
As $J_2$ increases, we observe three phases with distinct dimerized valence bond orders, which show an intricate competition with 
vector chiral order and incommensurate correlations.
For large $J_2$, regions of phase coexistence between the dimerized and vector chiral orders emerge.
We present large-scale numerical data for the determination of transition lines, order parameters, and the nature of the phase transitions.
\end{abstract}

\maketitle

\section{Introduction}\label{sec:intro}
Historically antiferromagnetic Heisenberg chains (AFHCs) have been a great inspiration for the development of 
theoretical and numerical progress for quantum many-body systems.
Famous theoretical milestones include the Bethe ansatz~\cite{Bethe1931, Karbach_1998}, bosonization~\cite{Luther1975}, and the discovery of symmetry protected topological (SPT) states with non-local string orders~\cite{haldane_1983_464_Continuum, haldane_1983_1153_Nonlinear, den_Nijs_1989, Gu_2009, Pollmann_2012}.
Remarkable numerical efforts on AFHCs date back to the early days of computers~\cite{Bonner1964} and have contributed to the advent of powerful tools like the density matrix renormalization group (DMRG)~\cite{white_1992_2863_Density, peschel_1999__DensityMatrix}.

By now it is well understood that the ground state for half-odd-integer chains is gapless with a power-law decay of correlations described by bosonization and the Wess-Zumino-Witten (WZW) SU($2$)$_1$ model of two dimensional conformal field theory (CFT)~\cite{affleck_1987_5291_Critical, hallberg_1996_4955_Critical}, while integer spin chains are gapped with a different ground state topology~\cite{haldane_1983_464_Continuum,haldane_1983_1153_Nonlinear}.
The situation becomes more interesting when frustration in the form of a nearest-neighbor (NN) coupling $J_1$ and a next-nearest-neighbor (NNN) coupling $J_2$ is introduced,
which induces a gap for $J_2 / J_1 \gtrsim 0.2411$~\cite{okamoto_1992_433_Fluiddimer, eggert_1996_R9612_Numerical} in the spin-$1/2$ AFHC.
Incommensurate correlations are observed~\cite{bursill_1995_8605_Numerical, white-affleck96, roth_1998_9264_Frustrated} as the ratio of the two couplings exceeds the Majumdar-Ghosh (MG) point $J_2/J_1 = 1/2$~\cite{Majumdar_1969_1, Majumdar_1969_2}.
However, for spin-$1$ AFHCs, the interplay between the $J_1$ and $J_2$-terms only results in a transition from the Haldane phase to the topologically trivial NNN Haldane phase, without the emergence of a dimerized phase~\cite{Kolezhuk_1996}.
Meanwhile, many studies on the spin-$1$ bilinear-biquadratic model~\cite{Chubukov_1990, Chubukov_1991, Rizzi_2005, Buchta_2005, Hu_2014} have been conducted, with findings of unconventional mechanisms of spontaneous dimerization~\cite{Rizzi_2005, Buchta_2005, Hu_2014}.
Distinct dimerized patterns, e.g., columnar dimer, staggered dimer, and valence bond solid (VBS) phases, have been observed in two-leg spin-$1/2$ frustrated $XXZ$ ladders with or without the ring exchanges~\cite{Metavitsiadis_2017, Ogino_2022}, a honeycomb ladder~\cite{Luo_2018}, the three-leg cases~\cite{Almeida_2008}, other two-dimensional systems~\cite{Starykh_2007, Gong_2016}, or at the places adhering to promising spin liquids~\cite{Metavitsiadis_2014, Wang_2018}.  Frustration in spin-ladder models may also lead to incommensurate correlations~\cite{nersesyan_1998_910_Incommensurate, Lavarelo2011, metav2017}.

 Helimagnetism with a so-called vector chiral order and broken space-inversion symmetry is only possible in finite magnetic fields or in the presence of a Dzyaloshinskii-Moriya (DM) interaction for the spin-1/2 AFHC~\cite{Zhao_2003, kolezhuk_2005_094424_Fieldinduced, mcculloch_2008_094404_Vector, chen_2011_094433_Spiral, guo_2023_045103_Spin}.
For higher spin systems, vector chiral order has been shown with frustrated anisotropic exchange couplings~\cite{lecheminant_2001_174426_Phase, hikihara_2001_174430_Groundstate,hikihara_ground-state_2002}.  A corresponding bosonization analysis was also done for both spin-$1/2$ and $1$, validated by numerical evidence~\cite{mcculloch_2008_094404_Vector}.

We now focus on the specific case of the spin-$3/2$ chain, which allows intricate interactions and different spin orders.
Existing studies have shown that a spin-$3/2$ can be regarded as a combination of a spin-$1/2$ and a spin-$1$, e.g., for the studies of the partially magnetic VBS and the floating phase~\cite{oshikawa_magnetization_1997,chepiga_2020_174407_Floating}.
For the spin-$3/2$ $J_1$-$J_2$ chain, frustration causes spontaneous dimerization when $J_2 / J_1 > 0.29$, with edge states playing significant roles~\cite{roth_1998_9264_Frustrated,ng_1994_555_Edge,qin_1995_12844_Edge}.
Recent studies have further elucidated that various dimerized phases, as well as an incommensurate and gapless floating phase emerge when three-spin interactions are considered in arrays of Rydberg atoms~\cite{chepiga_2020_174407_Floating}.
More excitingly, the floating phase has been detected in a recent experiment by looking at site-resolved Rydberg state densities ~\cite{Zhang_2024}.
A widely observed uniaxial single-ion anisotropy $D$-term resulting from crystal field splitting is given by the quadrupolar operator along the $z$-axis.
Such a term is trivial for spin-$1/2$, but for the spin-$3/2$ AFHC it was found that a gapped state exists for $0.387 < D / J_1< 0.943$ in finite magnetic fields, leading to a fractional $1/3$-magnetization plateau~\cite{oshikawa_1997_1984_Magnetization, sakai_1998_R3201_Magnetization, kitazawa_2000_940_Magnetizationplateau}.
Without magnetic field the spin-$3/2$ chain is in a gapless state for $D>0$.

The combined effect of single-ion anisotropy $D>0$ and frustration $J_2$ is so far unclear.
We now present the corresponding detailed groundstate phase diagram of the spin-$3/2$ $J_1$-$J_2$-$D$ model in zero magnetic field, which displays an interesting competition between dimerization and vector chiral order.
The paper is organized as follows:
We first introduce the model and phase diagram in Sec.~\ref{sec:modelandphasediagram},
then proceed to characterize dimerized phases and discuss phase transitions in Sec.~\ref{sec:dimerizedphases}.
Next, we demonstrate the emergence of vector chirality, investigate the instability of the floating phase, and give the boundary lines of phase coexistence regions in Sec.~\ref{sec:vectorchirality}.
Finally, we present our concluding remarks in Sec.~\ref{sec:conclusion}.

\begin{figure}[t]
\centering
\includegraphics[width=\columnwidth]{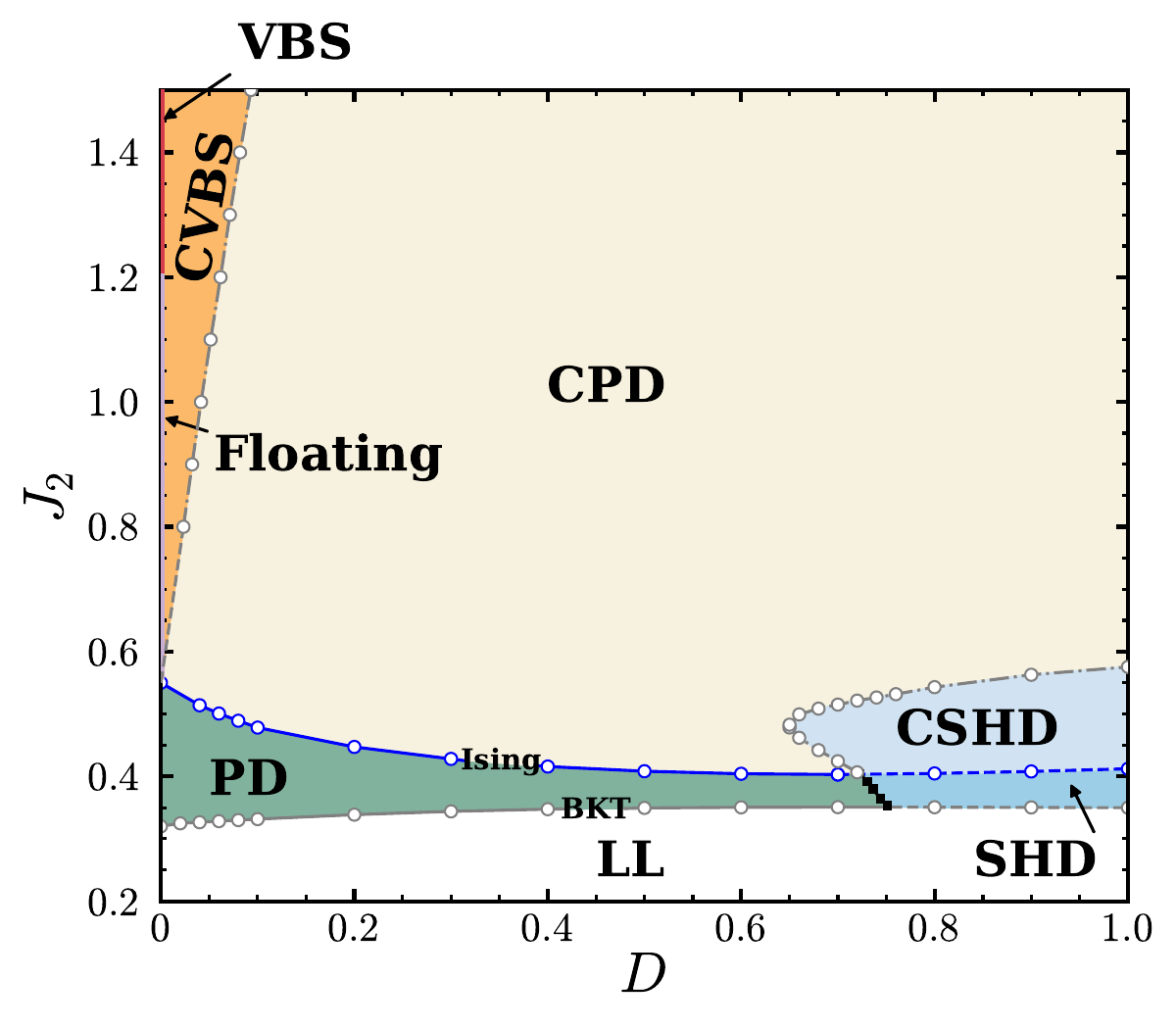}
\caption{$D$-$J_2$ Groundstate phase diagram, consisting of a Luttinger liquid (LL) phase, a floating phase, a partially dimerized (PD) phase, a spin-$1/2$-like dimerized (SHD) phase, a valence bond solid (VBS) phase, and three phase coexistence regions, including a chiral PD (CPD) phase, a chiral SHD (CSHD) phase, and a chiral VBS (CVBS) phase.
A LL-PD (gray solid line) and a LL-SHD (gray dashed line) transitions of BKT type, a PD-CPD (blue solid line) and an SHD-CSHD (blue dashed line) transitions of Ising type, a PD-SHD (black solid line) crossover, and CPD-CVBS and CPD-CSHD boundaries (gray dash-dotted line) indicated by zero dimerization are plotted.
Transition points ({\color{gray}$\circ$}, \tcb{$\circ$} and $\blacksquare$) are marked.
\label{fig:Phasediagram}}
\end{figure}

\section{Model and phase diagram}\label{sec:modelandphasediagram}

We consider a Hamiltonian given by
\begin{equation}
{\cal H} = \sum^L_{\ell=1} \left[ J_1 \left(\mathbf{S}_\ell \cdot \mathbf{S}_{\ell+1}\right) + J_2 \left(\mathbf{S}_{\ell} \cdot \mathbf{S}_{\ell+2}\right) + D \left(S_\ell^z\right)^2\right]\, ,
\label{eq:Hamiltonian}
\end{equation}
where the operator $\mathbf{S}_\ell$ represents the spin-$3/2$ at site-$\ell$ and comprises of components $S^x_\ell$, $S^y_\ell$ and $S^z_\ell$ in the $x$, $y$ and $z$-axes.
$J_1$ and $J_2$ are the coupling strengths of the NN and NNN Heisenberg interactions, respectively.
In the following, we set $J_1 = 1$ as the energy unit.
In the case of ferromagnetic coupling $J_2 < 0$, there is no competition between the $J_1$ and $J_2$ terms in the model~\eqref{eq:Hamiltonian} and the ground state remains monospecific, similar to spin-$1/2$ and spin-$1$~\cite{bursill_1995_8605_Numerical, murashima_2005_1544_Phase}.
Thus, in this work we consider the more interesting region $J_2 \ge 0$.
The strength of the uniaxial single-ion anisotropy is given by $D\ge 0$.
The site index $\ell$ ranges from $1$ to the length of the chain $L=2N$, and $N$ is an integer.
In our convention, the left terminal point on the even bond-$2n$ is located at site-$2n$, with $1 \le n \le N$.

The groundstate phase diagram~Fig.~\ref{fig:Phasediagram} with zero magnetization $S^z_\text{t} = \sum^L_{\ell=1} S^z_\ell = 0$ is known for special cases:
(\textbf{i}) For a spin-$3/2$ Heisenberg chain lacking both $D$ and $J_2$ terms, the ground state exhibits critical behavior described by the WZW$_{k=1}$ theory with a central charge $c=1$~\cite{hallberg_1996_4955_Critical}, analogous to the case of a spin-$1/2$ Heisenberg chain.
(\textbf{ii}) In the case of $D > 0$ and $J_2=0$, the ground state remains in the Luttinger liquid (LL) phase.
(\textbf{iii}) In the limit of $D =+\infty$, the states with $S^z_\ell = \pm 3/2$ are forbidden for all sites in the ground state, thus rendering the single-ion anisotropy a trivial identity operator.
Consequently, the model~\eqref{eq:Hamiltonian} can be effectively mapped to a spin-$1/2$ $J_1$-$J_2$ $XXZ$ model with the anisotropy parameter $\Delta_\text{eff} = 0.25 < 1$, which has been extensively studied previously~\cite{nomura_1994_5773_Critical}.
This model undergoes a BKT-type quantum phase transition from an LL phase to a spin-$1/2$ dimerized phase (SHD), by breaking the discrete translational symmetry~\cite{eggert_1996_R9612_Numerical}.
(\textbf{iv}) Along the $D=0$ line, as $J_2/J_1$ increases, system changes from the LL phase to a partially dimerized (PD) phase, and then to the floating phase~\cite{chepiga_2020_174407_Floating}. In the dominant-$J_2$ limit, the ground state favors the VBS phase, rather than a fully dimerized (FD) phase previously proposed~\cite{chepiga_2020_174407_Floating}, which will be clarified later in Sec.~\ref{subsec:VBSvsFD}.

In this work, we use the exact diagonalization (ED), the density matrix renormalization group (DMRG)~\cite{white_1992_2863_Density,peschel_1999__DensityMatrix}, and the infinite DMRG (iDMRG)~\cite{mcculloch_2008__Infinite, hu_2011_220402_Accurate} methods to determine the details of the groundstate phase diagram Fig.~\ref{fig:Phasediagram}, which can be outlined as follows:
(1) We identify three distinct dimerized phases: a PD phase~\cite{Rachel_2009}, a SHD phase, and a VBS phase.
(2) The interplay between dimerization and vector chirality leads to the emergence of phase coexistence in a chiral PD (CPD) phase, a chiral SHD (CSHD) phase, and a chiral VBS (CVBS) phase.
(3) Both the floating phase and the VBS phase exhibit an instability toward single-ion anisotropy, which exists only in a narrow region near the $J_2$ axis.
(4) Complicated phase transitions occur between these phases.
These results will be presented in the following sections.

\section{Dimerized phases}\label{sec:dimerizedphases}

\begin{figure}[t]
\centering
\includegraphics[width=\columnwidth]{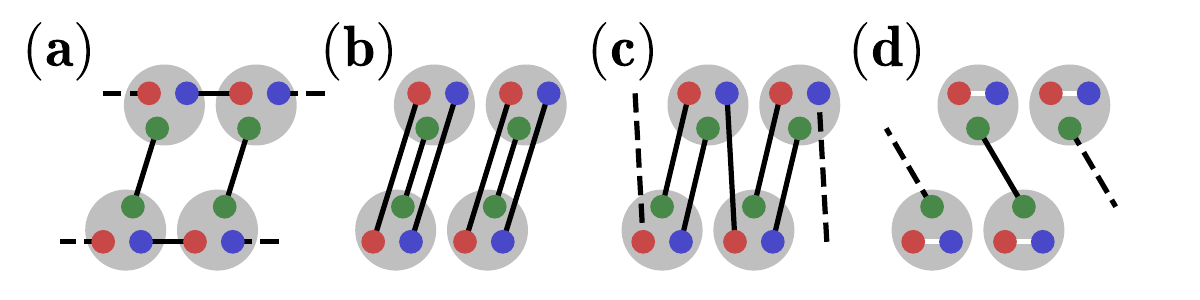}
\caption{(Color online)
Mean-field ansatz in the valence bond representation:
(a) VBS with NN and NNN spin-$1/2$s are equally linked.
(b) The fully dimerized (FD) phase~\cite{Michaud_2012} with three VBS singlets on every second NN bond.
(c) PD with one and two VBS singlets on the even and odd bonds, respectively.
(d) SHD, where two of the three spin-$1/2$s are supressed by $D$, so that only one VBS singlet is present on every second NN bond.
A red, green or blue circle represents one spin-$1/2$, and a black line represents a VBS singlet of two spin-$1/2$s.
A gray shadow denotes a fully symmetric projection of three spin-$1/2$s.
\label{fig:dimer_vbs}}
\end{figure}

Dimerized phases have finite dimerization strength $\lvert \mathcal{D} \rvert$, defined as the strength discrepancy of even and odd NN bonds in the middle of the chain in the thermodynamic limit (TDL)
\begin{equation}
\mathcal{D} = \lim_{N\rightarrow+\infty} \braket{\left( \mathbf{S}_{N - 1} - \mathbf{S}_{N+1} \right) \cdot \mathbf{S}_N}\, ,
\end{equation}
which is also called the dimerization~\cite{Furukawa_2012, Ejima_2018, Xu_2021}.
In Fig.~\ref{fig:dimer_vbs}, each dimerized phase has a mean-field ansatz for the bulk in the valence bond representation, in which each spin-$3/2$ can effectively be mapped to three spin-$1/2$s, colored by $c= \text{r}$ (red), $\text{g}$ (green), and $\text{b}$ (blue), respectively.
In detail, the basis $\ket{S^z_\ell}$ for a spin-$3/2$ at site-$\ell$ can be spanned as 
\begin{eqnarray}
\ket{S^z_\ell} = \sum_{\sigma^\text{r}_\ell,\ \sigma^\text{g}_\ell,\ \sigma^\text{b}_\ell = \uparrow, \downarrow} \mathcal{P}_\ell (\sigma^\text{r}_\ell, \sigma^\text{g}_\ell, \sigma^\text{b}_\ell; S^z_\ell) \ket{\sigma^\text{r}_\ell} \otimes \ket{\sigma^\text{g}_\ell} \otimes \ket{\sigma^\text{b}_\ell}\, ,
\end{eqnarray}
where $\sigma^{\text{r}/\text{g}/\text{b}}_\ell = \uparrow$,$\downarrow$ mark the $z$-axis spin polarization of the spin-$1/2$, and the uniform parameter $\mathcal{P}_\ell \equiv \mathcal{P}$ gives the fully symmetric projection operator of three spin-$1/2$s.
For example, following Fig.~\ref{fig:dimer_vbs}(a), we can write down the ansatz wave function for VBS, i.e.,
\begin{equation}
\ket{\psi_\text{VBS}} = \left(\otimes_\ell \mathcal{P}_\ell\right) \left[\cdots \ket{\phi^{\text{g},\text{g}}_{1,2}} \ket{\phi^{\text{b},\text{r}}_{1,3}} \ket{\phi^{\text{b},\text{r}}_{2,4}} \ket{\phi^{\text{g},\text{g}}_{3,4}} \cdots\right]\, ,
\end{equation}
and the state $\ket{\phi^{c,c'}_{\ell,\ell'}} = \left(\ket{\uparrow_{\ell,c} \downarrow_{\ell',c'}} - \ket{\downarrow_{\ell,c} \uparrow_{\ell',c'}}\right) / \sqrt{2}$ denotes a singlet between spin-$1/2$-$c$ at site-$\ell$ and spin-$1/2$-$c'$ at site-$\ell'$.
It should be noted that no internal valence bond states between two spin-$1/2$s on the same site are possible.
However, at large $D$ the $S^z = \pm 3/2$ states are forbidden, which effectively freezes out two of the three spin-$1/2$ in that limit as depicted by white internal lines in Fig.~\ref{fig:dimer_vbs}(d).

\subsection{VBS vs. FD}\label{subsec:VBSvsFD}
\begin{figure}[t]
\centering
\includegraphics[width=\columnwidth]{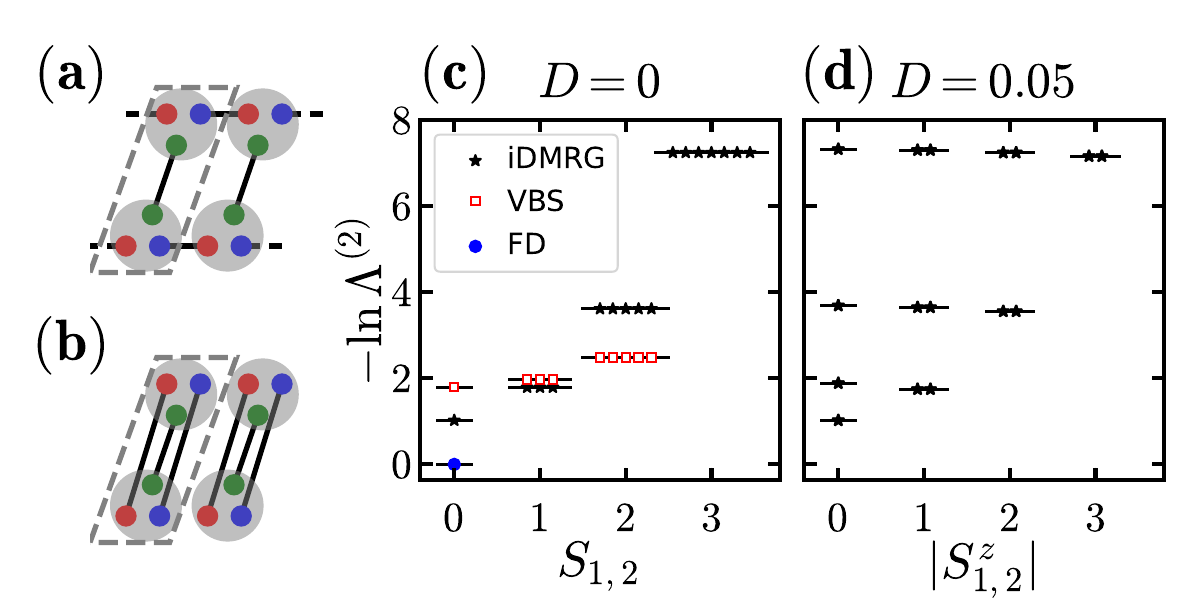}
\caption{(Color online)
After obtaining the two-site (enclosed by a gray quadrangle) reduced density matrix $\rho^{(2)}$ defined in Eq.~\eqref{eq:rho12} for (a) VBS and (b) FD by iDMRG, the entanglement spectra $\{-\ln \Lambda^{(2)}_\alpha \}$ are plotted as a function of (c) the total spin $S_{1,2}$ at $D=0$, and (d) the $z$-axis spin polarization $S^z_{1,2}$ at $D=0.05$.
In (c), the entanglement spectrum of the mean-field ansatz is plotted for comparison.
$J_2=4$ and the truncated bond dimension $m=3,200$ are used.
\label{fig:vbs_fd_spectrum}}
\end{figure}

In the $J_2 = +\infty$ limit of the $D=0$ line, either the VBS phase or the FD phase, which reflect different internal symmetries of spin-halves within a spin-$3/2$, can be chosen as a candidate for the ground state.
Nevertheless, distinguishing between these two phases remains a significant challenge, as they typically exhibit differences in short-range correlations and does not cause any phase transition usually, such as in the groundstate phase diagram of the spin-$3/2$ $J_1$-$J_2$-$J_3$ model~\cite{chepiga_2020_174407_Floating}.
In VBS, each of two decoupled spin-$3/2$ Heisenberg chains can be seen as a spin-$1$ chain with one dangling spin-$1/2$ per site.
In iDMRG calculations, we can obtain both of the twofold degenerate ground states of the VBS phase and the FD phase by appropriately setting the initial conditions~\cite{mcculloch_2008__Infinite, hu_2011_220402_Accurate}.
In one of degenerate manifolds for the VBS phase, as shown in Fig.~\ref{fig:vbs_fd_spectrum}(a),
due to the perturbative $J_1$ term, two dangling spin-$1/2$s are connected to form a VBS singlet at each odd NN bond.
In contrast, the FD phase has three VBS singlets at every other NN bond.
Since both phases have finite $\lvert\mathcal{D}\rvert$ and the similar low-lying multiplet of the entanglement spectrum for the bipartite reduced density matrix~\cite{pollmann_2010_064439_Entanglement}, it is difficult to distinguish them.
Alternatively, we calculate the reduced density matrices
\begin{equation}\label{eq:rho12}
\rho^{(2)} = \text{tr}_{\overline{1,2}} \ketbra{\psi}{\psi}
\end{equation}
for sites $1$ and $2$ on the odd bond, after tracing out other degrees of freedom $\overline{1,2}$ in the groundstate wave function $\ket{\psi}$.
In Fig.~\ref{fig:vbs_fd_spectrum}(c), we first show the entanglement spectrum $\{-\ln \Lambda^{(2)}_\alpha \}$ of the mean-field ansatz for VBS and FD, and $\Lambda^{(2)}_\alpha$ gives the $\alpha$-th eigenvalues of $\rho^{(2)}$ with $\alpha=1$, $\cdots$, $16$.
The FD phase has a unique $\Lambda^{(2)}_1 = 1$ or equivalently $\ln \Lambda^{(2)}_1 = 0$ in the spectrum, while the VBS phase has a structure of ``1+3+5" corresponding to the levels with the total spin $S_{1,2} = \lvert \mathbf{S}_1 + \mathbf{S}_2 \rvert = 0$, $1$, and $2$, respectively.
In comparison with the iDMRG result at $D=0$ and $J_2=4$, shown in Fig.~\ref{fig:vbs_fd_spectrum}(c), we find that the multiplet structure of $\rho_{1,2}$ shows behavior more like the mean-field VBS ansatz than the FD one.
In addition, we present the result at $D=0.05$, where the SU($2$) symmetry is broken, leading to a spectrum as a function of the total $z$-axis spin polarization $S^z_{1,2} = S^z_1 + S^z_2$.

\subsection{PD vs. SHD}\label{subsec:PDvsSHD}

\begin{figure}[t]
\centering
\includegraphics[width=\columnwidth]{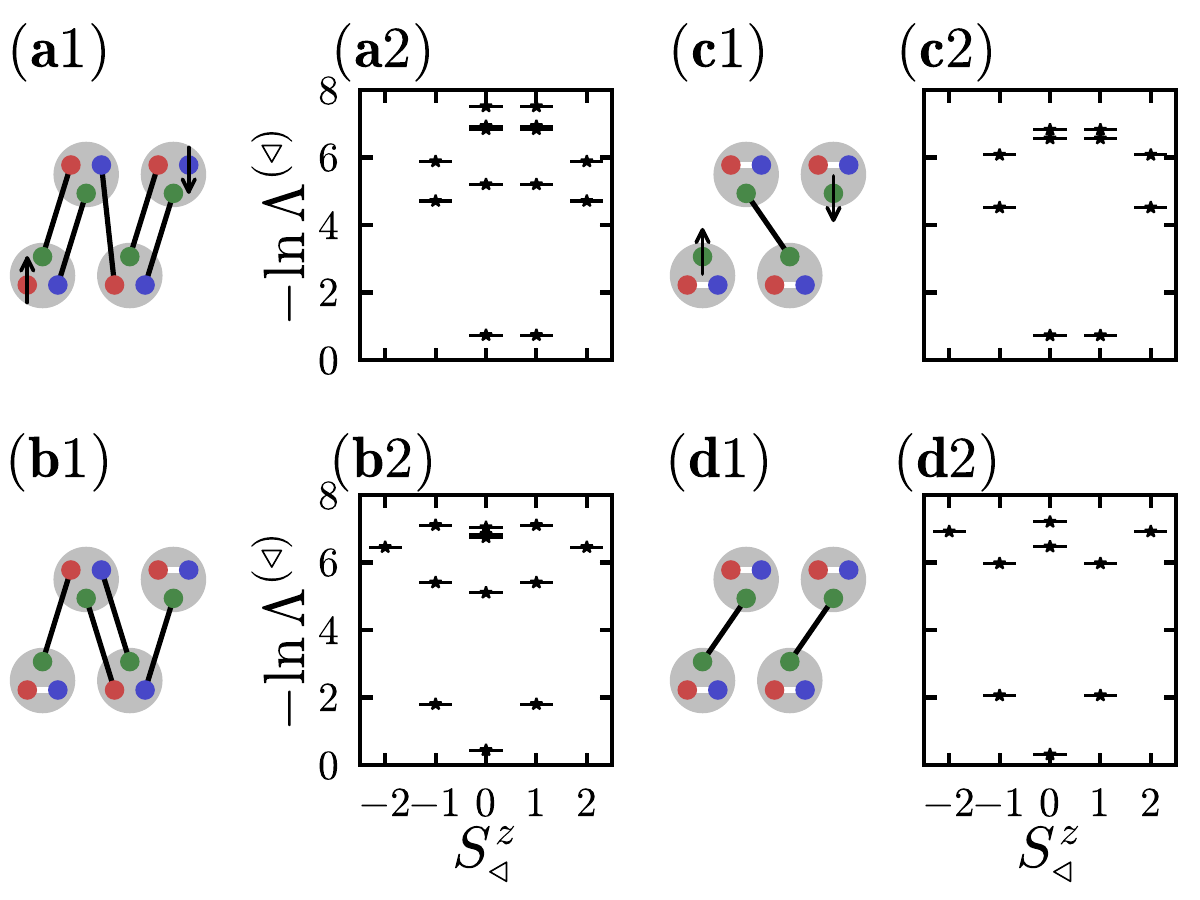}
\caption{(Color online)
Mean-field ansatz in the valence bond representation for (a1) a PD edge state, (b1) a PD no-edge state, (c1) a SHD edge state, and (d1) a SHD no-edge state, under OBC.
In the case of even $N$, the entanglement spectra $\{-\ln \Lambda^{(\triangleleft)}_\alpha\}$ for the middle even bond in (a2) the PD edge state and (b2) the PD no-edge state at $D=0.6$, and (c2) the SHD edge state and (d2) the SHD no-edge state at $D=0.8$, are calculated by iDMRG.
$J_2=0.403$ and $m=3,200$ are used in iDMRG.
\label{fig:PDSHD}
}
\end{figure}

The edge states in the dimerized phases also involve an additional degree of degeneracy in the ground state, which have led to different estimates of the phase boundary of PD at $D=0$~\cite{roth_1998_9264_Frustrated, chepiga_2020_174407_Floating}.
One identified the disappearance of the edge states as a signal of a $1$st-order phase transition~\cite{roth_1998_9264_Frustrated}, while the latter proposed a mechanism of reorientation of dimers resulting in a different phase boundary~\cite{chepiga_2020_174407_Floating}.
Under the periodic boundary condition (PBC), the PD phase characterized by the presence of doubly degenerate ground states, one of which exhibits either strong even or odd bonds.
However, under the open boundary condition (OBC), two edge spin-$1/2$s are coupled to form a singlet for a short chain.
This occurs when the odd bond consists of two VBS singlets and is stronger than the even bond in the mean-field ansatz.
The spins are disentangled when their distance $L-1$ is much larger than the correlation length~\cite{chepiga_2020_174407_Floating}, yielding spin polarizations $\uparrow$ and $\downarrow$ located at two edges, respectively, as shown in Fig.~\ref{fig:PDSHD}(a1), which is called the ``PD edge state".
By simultaneously flipping these two spin-$1/2$s, another degenerate ground state can be obtained in the $S^z_\text{t}=0$ sector.
In contrast, when the even bonds are stronger, the ground state has no edge spin-$1/2$s, as shown in Fig.~\ref{fig:PDSHD}(b1), called the ``PD no-edge state".

With iDMRG using distinct initial conditions~\cite{mcculloch_2008__Infinite, hu_2011_220402_Accurate}, we can capture all degenerate patterns, characterized by the distinct entanglement spectrum $\{-\ln \Lambda^{(\triangleleft)}_\alpha\}$ of the reduced density matrix for the left semi-chain ($\triangleleft$) $\rho^{(\triangleleft)} = \text{tr}_\triangleright \ketbra{\psi}{\psi}$, when we divide the chain into two equal semi-chains at the middle even bond.
These patterns include two PD edge states (Fig.~\ref{fig:PDSHD}(a1) for one manifold) and a unique PD no-edge state (Fig.~\ref{fig:PDSHD}(b1)).
Fig.~\ref{fig:PDSHD}(a2) illustrates a doubly degenerate and symmetric spectrum with respect to the $z$-axis spin polarization $S^z_\triangleleft = 1/2$ for $\triangleleft$.
This indicates that the central bond has a single VBS singlet, and a spin-$1/2$ is polarized up and localized at the left edge, consistent with the scenario for the PD edge state shown in Fig.~\ref{fig:PDSHD}(a1).
In contrast, the twofold degeneracy of the spectrum is absent in Fig.~\ref{fig:PDSHD}(b2) and is symmetric with respect to $S^z_\triangleleft = 0$, which aligns with the image for the PD no-edge state shown in Fig.~\ref{fig:PDSHD}(b1).
Moreover, we use the white internal valence bonds for representing its complex edge spin patterns, rather than effective spin-$1/2$s localized at edges.
Compared to previous work on finite chains at $D=0$, twofold degenerate PD edge states should correspond to the ground states, carrying spin-$1/2$ edge states as well, in the combined regions of ``PDC" and ``PDIC"~\cite{chepiga_2020_174407_Floating}.
While the unique PD no-edge state may be analogous to the one in the ``RPD" region~\cite{chepiga_2020_174407_Floating}.
By increasing $D$, the bases with $S^z_\ell = \pm 3/2$ at all lattice sites are suppressed, each spin-$3/2$ becomes an effective spin-$1/2$, with two spin-$1/2$s form a VBS singlet at each lattice site, as shown in Fig.~\ref{fig:PDSHD}(c1,d1).
Fig.~\ref{fig:PDSHD}(c2) and (d2) show the entanglement spectrum for the SHD edge and no-edge states, respectively.

\subsection{PD-SHD transition}\label{subsec:PDvsSHD}

We expect that the uniaxial single-ion anisotropy may facilitate a channel for connecting the PD and SHD edge states at the PD-SHD transition point.
Similarly, a channel may also be formed between two no-edge states.
In the iDMRG calculation, the symmetric center of the entanglement spectrum can be employed to characterize the emergence of the edge states.
\begin{figure}[t]
\centering
\includegraphics[width=\columnwidth]{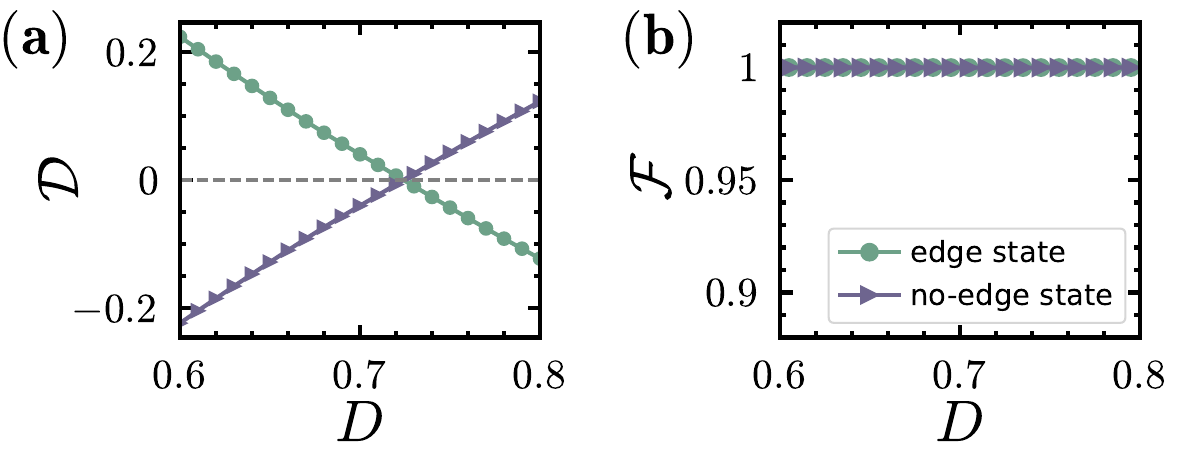}
\caption{(Color online) (a) The dimerization $\mathcal{D}$ at even $N$ and (b) the fidelity $\mathcal{F}$ for the edge and no-edge states along a cutting line $J_2=0.403$ are calculated by iDMRG.
In (a), $\mathcal{D}$ curves intersect at the zero-dimerization axis, indicating the PD-SHD transition point.
$m=3,200$ is used in the calculation.
\label{fig:first}
}
\end{figure}
In Fig.~\ref{fig:first}(a), we find that the dimerization $\mathcal{D}$ for both the edge and no-edge states change the sign at the same place $D^\text{1\text{st}}_\text{c} \approx 0.724$.
Meanwhile, we monitor the fidelity
\begin{equation}\label{eq:fidelity}
\mathcal{F}= \vert \braket{\psi(D) \vert \psi(D+\delta D)} \vert
\end{equation}
in both the channels for the edge and no-edge states, where we choose the interval $\delta D = 0.01$.
As illustrated in Fig.~\ref{fig:first}(b), we find that $\mathcal{F}$ is nearly one with a discrepancy of less than $10^{-5}$, which means that the PD phase can undergo an adiabatic transition to the SHD phase, despite the violation of two VBS singlets at arbitrary odd bonds.
Besides, we find no signal of the singularity in correlation lengths in the vicinity of the transition point (not shown).
Therefore, we categorize this as a crossover~\cite{tonegawa_haldane_2011, pollmann_symmetry_2012}.
However, unlike traditional crossovers that involve unique ground states, this crossover from the PD phase to the SHD phase happens between multi-fold degenerate ground states.
In Fig.~\ref{fig:Phasediagram}, five crossover points are indicated by black squares and situated at the coordinates $(D,\, J_2)=(0.723,\, 0.404)$, $(0.73,\, 0.392)$, $(0.736,\, 0.38)$, $(0.744,\, 0.364)$, and $(0.751,\, 0.353)$ from up to bottom, respectively.

\subsection{LL-PD and LL-SHD transitions}\label{subsec:bkt}

\begin{figure}[b]
\centering
\includegraphics[width=\columnwidth]{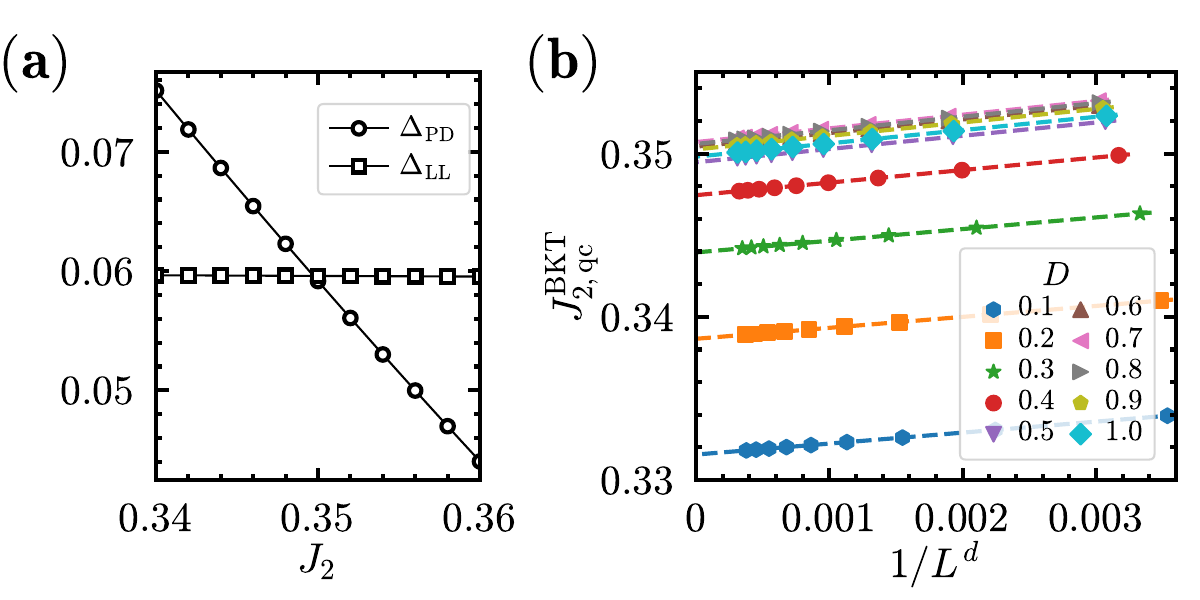}
\caption{(Color online)
(a) The determination of the quasicritical point $J^\text{BKT}_{2,\text{qc}}$ at $D=0.5$ and $L=48$, indicated by the intersection point of the gap curves for two representative excited energy levels $\Delta_\text{LL}$ and $\Delta_\text{PD}$ under PBC.
(b) A power-law fit of the quasicritical points $J^\text{BKT}_{2,\text{qc}}$ to the chain length $L$ for different values of $D$.
We use ED for the case $L < 20$, while we use DMRG with $m=4,096$ when $20 \le L \le 48$.
\label{fig:BKT}
}
\end{figure}

It has been established that the transition from the LL phase to either the PD phase at $D=0$~\cite{chepiga_2020_174407_Floating}, or the SHD phase in the limit of $D=+\infty$ is known to be of BKT type, as described by the WZW$_{k=1}$ theory~\cite{affleck_1987_5291_Critical, chepiga_2020_174407_Floating}.
When $D > 0$, there is no extra relevant mass term in the presence of the low-energy effective theory describing the LL-PD transition, which means that it can remain of BKT type.

To verify this, we employ the level-spectroscopy method~\cite{okamoto_1992_433_Fluiddimer, nomura_1994_5773_Critical, nomura_1995_5451_Correlation, zhong_2022_045130_Quantum} to determine the place of the transition point $J^\text{BKT}_{2,\text{c}}$, based on symmetry and renormalization analysis.
It is assumed that the energy levels under the PBC are arranged in ascending order, and that the $s$-th (the index $s=0$, $\cdots$) level $E_s (S^z_\text{t},\, k,\, P)$ can be labeled by the quantum numbers of the total $z$-axis spin polarization $S^z_\text{t}$, the momentum $k$, and the parity $P$.
For even $N$, the ground state lies in the Hilbert space of $S^z_\text{t} = 0$, $k = 0$, and $P=+1$ in all relevant phases, so the corresponding energy curve is given by $E_\text{g} = E_0 (0,\, 0,\, +)$.
In this case, we choose the excited energy level $E_\text{LL} \equiv E_0 (+1,\, \pi,\, -)$ as the representative state for LL due to the spin-flipping symmetry, while we choose $E_{\text{PD}/\text{SHD}} \equiv E_0 (0,\, \pi,\, +)$ as the one for both PD and SHD.
In contrast, for odd $N$, the momentum $k$ is shifted by $\pi$ for all eigenstates, and the corresponding party $P$ is reversed either.

In Fig.~\ref{fig:BKT}, we use the ED and DMRG methods to compute $E_\text{g}$, $E_\text{LL}$ and $E_{\text{PD}/\text{SHD}}$ as a function of $J_2$ when $D$ is fixed.
In the first step, we interpolate to find the quasicritical point $J^\text{BKT}_{2,\text{qc}}$, where two representative gap curves intersect, i.e., $\Delta_\text{LL} \left(J^\text{BKT}_{2,\text{qc}}\right) = \Delta_{\text{PD}/\text{SHD}} \left(J^\text{BKT}_{2,\text{qc}}\right)$, where $\Delta_\text{LL} \equiv E_\text{LL} - E_\text{g}$ and $\Delta_{\text{PD}/\text{SHD}} \equiv E_{\text{PD}/\text{SHD}} - E_\text{g}$.
For example, for $L=2N=48$ and $D=0.5$, we get $J^\text{BKT}_{2,\text{qc}} \approx 0.352$ following a cubic interpolation, as shown in Fig.~\ref{fig:BKT}(a).
It is worth noting that the low-lying excited energy levels can be labelled using the quantum numbers in DMRG for large $L$, if we employ the guiding information provided by the evolution tendency of the levels obtained from ED for small values of $L$, such as the levels $E_\text{LL}$ and $E_{\text{PD}/\text{SHD}}$.
In the next step, we use a power-law function of $1/L^d$ to do data extrapolation to get the transition point $J^\text{BKT}_{2,\text{c}}$ in the TDL, as shown in Fig.~\ref{fig:BKT}(b), where the exponent $d$ is summarized in table~\ref{tab:exponent}.
\begin{table}[h!]
\centering
\begin{tabular}{c|cccccccc}
\hline
$D$ & $0$ & $0.1$ & $0.2$ & $0.3$ & $0.4$ & $0.5$ \\
\hline
$d$ & $2.046$\footnote{When $D=0$, the occasional degeneracy in the triplet states $S^z_\text{t}=0$, $\pm 1$ brings a trouble in determining $J^\text{BKT}_{2,\text{qc}}$.} & $2.036$ & $2.041$ & $2.058$ & $2.075$ & $2.087$  \\
\hline
\hline
$D$ & $0.6$ & $0.7$ & $0.8$ & $0.9$ & $1$ \\
\hline
$d$ & $2.092$ & $2.093$ & $2.092$ & $2.089$ & $2.086$ \\
\hline
\end{tabular}
\caption{Values of the exponent $d$ in the best fitting to $J^\text{BKT}_{2,\text{qc}}$.}
\label{tab:exponent}
\end{table}

We can see that $d$ is very close to $2$, which is equal to the exponent for the BKT-type transition in the spin-$1/2$ $J_1$-$J_2$ chain~\cite{okamoto_1992_433_Fluiddimer, tzeng_2012_024403_Parity}. This suggests that two transition may be subsumed within a unified low-energy theoretical framework.
Moreover, the maximum discrepancy between the $d$ value and $2$ implies a strong finite-size effect at $D=0.7$, which may be attributed to the competition between PD and SHD near the crossover.

\section{Vector chirality}\label{sec:vectorchirality}

\subsection{Phase coexistence}\label{subsec:phasecoexistence}

\begin{figure}[t]
\centering
\includegraphics[width=\columnwidth]{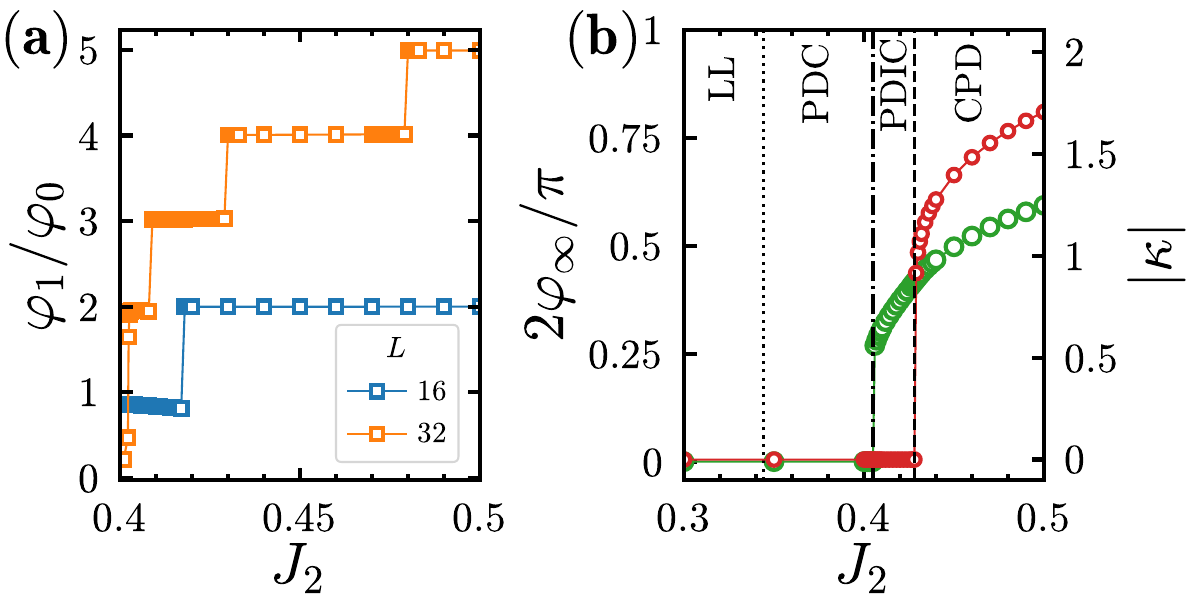}
\caption{(Color online)
(a) Flux $\varphi_1 / \varphi_0$ as a function of $J_2$ for $L=16$ and $32$ at $D=0.7$ is calculated by DMRG with $m=1,024$ under PBC.
(b) Flux $\varphi_\infty$ (green) and vector chirality $\lvert \kappa \rvert$ (red) as a function of $J_2$ at $D=0.3$ is calculated by iDMRG with $m=800$ under OBC.
The dotted and dashed lines indicate the LL-PD transition point $J^\text{LL-PD}_{2,\text{c}} \approx 0.3439$ and the PD-CPD transition point $J^\text{PD-CPD}_{2,\text{c}} \approx 0.428$, respectively.
The dash-dotted line indicates the place $J^\text{PDC-PDIC}_{2,\text{c}} \approx 0.405$ where a crossover between the PDC and PDIC regions occurs.
\label{fig:flux}
}
\end{figure}

When $J_2$ becomes sufficiently large, the space-inversion symmetry may be spontaneously broken in the ground state, resulting in a finite vector chirality
\begin{equation}\label{eq:vectorchiral}
\kappa = \lim_{N\rightarrow+\infty} \kappa_N = \lim_{N\rightarrow+\infty} \braket{\left( \mathbf{S}_N \times \mathbf{S}_{N+1} \right)_z}\, .
\end{equation}
where $\kappa_N = (i/2) [ \braket{S^+_N S^-_{N+1}} - \braket{S^-_N S^+_{N+1}} ]$ is the value defined in the middle NN bond.
For the vector chiral phase with $D>0$, we get $\braket{S^+_1 S^-_2} = \braket{S^+_N S^-_{N+1}} = \exp(-i \varphi_1)$, so that $\braket{S^-_1 S^+_2} = \braket{S^-_N S^+_{N+1}} = \exp(i \varphi_1)$ and $\kappa = \sin\varphi_1$ is governed by the flux $\varphi_1$~\cite{kolezhuk_2000_R6057_Quantum, dillenschneider_2008_732_Vector}.
In the limit of infinite distance $N = +\infty$, it can be shown that $\braket{S^+_1 S^-_2 S^+_N S^-_{N+1}} \approx \braket{S^+_1 S^-_2} \braket{S^+_N S^-_{N+1}} = \exp(-2 i \varphi_1)$ and $\varphi_1$ is determined by the equation
\begin{equation}\label{eq:fluxquanta}
\cos 2\varphi_1 \approx \frac{\braket{S^+_1 S^-_2 S^+_N S^-_{N+1}} + \braket{S^-_1 S^+_2 S^-_N S^+_{N+1}}}{\braket{S^+_1 S^-_2 S^-_N S^+_{N+1}} + \braket{S^-_1 S^+_2 S^-_N S^+_{N+1}}}\, . 
\end{equation}
For a chain with small $L$ less than the healing size for the vector chiral order, Eq.~\eqref{eq:fluxquanta} can still be used to derive the flux, although the ground state holds the space-inversion symmetry.
In the CPD region at $D=0.7$ in Fig.~\ref{fig:flux}(a), the vector chiral order emerges with $\varphi_1 / \varphi_0$ being the stepping function of $J_2$, which chooses integer numbers and $\varphi_0 = 2\pi / L$ represents the flux quanta.

In the long-range spin correlation $\lim_{x \rightarrow +\infty} \braket{S^+_N S^-_{N+x}} \propto \exp(-i \varphi_\infty x)$, we also monitor another flux
\begin{equation}\label{eq:correlationlength}
2 \varphi_\infty = \text{Arg} (\gamma_1 / \gamma_0)\, ,
\end{equation}
where $\gamma_0$ and $\gamma_1$ represent the dominant eigenvalues in the $S^z_\text{t} = 0$ and $+1$ sectors of the transfer matrix~\cite{mcculloch_2008__Infinite, he_2024_035126_Phase}, respectively.
A variational unit cell comprising two sites was used in the iDMRG method.
The flux $\varphi_\infty$ is equivalent to the pitch angle defined in the mean-field scenario~\cite{radcliffe_1971_313_properties}.
As illustrated in Fig.~\ref{fig:flux}(b) at $D=0.3$, $2 \varphi_\infty \ge 0$ modulo $2\pi$ is observed as $J_2$ increases and goes across the LL-PD and PD-CPD transition points successively.
$2 \varphi_\infty = 0$ consistent with $\varphi_\infty = \pi$ near the LL-PD transition corresponds to the antiferromagnetic spin correlation, while $2 \varphi_\infty \ne 0$ in the vicinity of the PD-CPD transition indicates an incommensurate spin correlation.
Thus, indicated by the short-range spin correlation in the PD phase region, the ground state undergoes a crossover from a commensurate PD (PDC) region to an incommensurate PD (PDIC) region at $J^\text{PDC-PDIC}_2 \approx 0.405$.

\begin{figure}[t]
\centering
\includegraphics[width=\columnwidth]{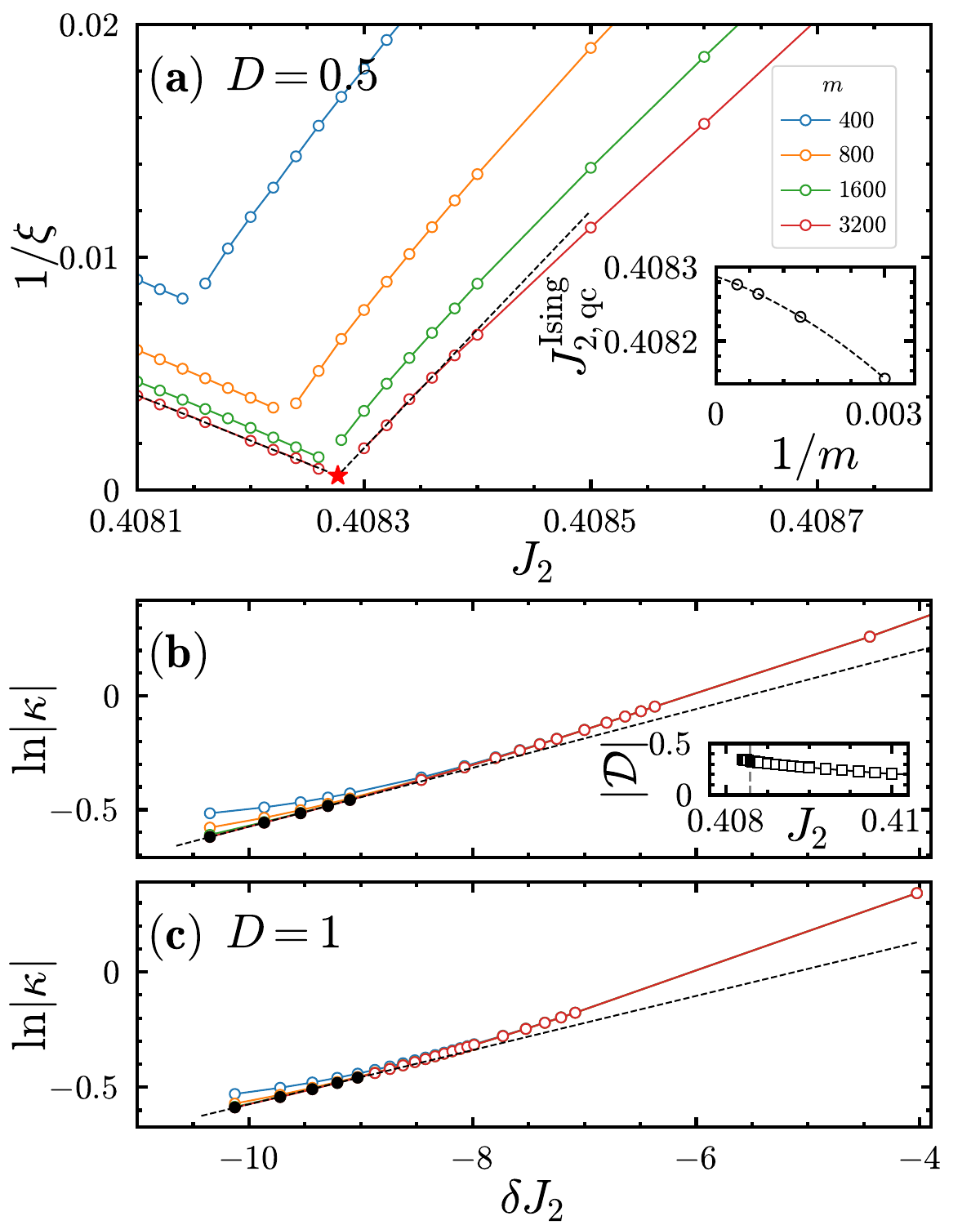}
\caption{(Color online)
(a) The inverse of the correlation length $\xi$ calculated by iDMRG as a function of $J_2$ for various $m$ at $D=0.5$.
The quasicritical point $J^\text{Ising}_{2,\text{qc}}$ (\tcr{$\star$} for $m=3,200$) is determined through a fitting procedure whereby $1/\xi \propto \left\lvert \delta J_2 \right\rvert^\nu$ near the minima, where $\nu$ is found to be very close to $1$.
Inset: $J^\text{Ising}_{2,\text{qc}}$ is plotted as a function of $m$ and the value $J^\text{Ising}_{2,\text{c}}$ (dashed-dotted line) in the $m=+\infty$ limit is obtained by fitting to the parabolic curve of $1/m$.
The vector chirality $\lvert\kappa\rvert$ as a function of $\lvert \delta J_2 \rvert$ in the log-log scale for various $m$ are plotted at (b) $D=0.5$ and (c) $1$.
In panels (b) and (c), the linear fitting (dashed line) to the data points ($\bullet$) suggests $\lvert \kappa \rvert \propto \lvert \delta J_2 \rvert^\beta$ with $\beta=0.129$ and $0.121$, respectively, both of which are close to $1/8$.
\label{fig:PD_CHIRAL}
}
\end{figure}

A comprehensive scenario can be delineated within the PD phase region.
At very shortest range, such as the NN and NNN bonds, the spin correlation is always antiferromagnetic in nature, exhibiting no vector chirality.
Nevertheless, at a somewhat longer distance, the spin correlation remains antiferromagnetic with $\varphi_\infty=\pi$, or alternatively, it may turn incommensurate with $\varphi_\infty \ne \pi$, as a consequence of the presence of the triplet domain walls.
The comparable phenomenon occurs in spin-$1/2$ AFHCs, where the interplay between short-range spin correlations gives rise to the discrepancy between the disordered point and the Liftshiftz point~\cite{bursill_1995_8605_Numerical, roth_1998_9264_Frustrated}.
When entering the CPD region with a nonzero vector chirality, an incommensurate spin correlation on the NN bond and in the long distance has been established, with $\varphi_1 \ne 0$.
Therefore, throughout the PD-CPD transition, the incommensurate spin order persists, and thus the PDIC region is certainly sandwiched by the PDC and CPD regions.

Next, we determine the criticality at the PD-CPD transition point where $\lvert\kappa\rvert$ emerges.
In Fig.~\ref{fig:PD_CHIRAL}, we calculate the correlation length, defined as
\begin{equation}\label{eq:correlationlength}
\xi = -2 a / \ln \left\lvert \gamma_2 / \gamma_0 \right\rvert\, ,
\end{equation}
where $\gamma_2$ represents the second dominant eigenvalue in the $S^z_\text{t} = 0$ sector of the transfer matrix~\cite{mcculloch_2008__Infinite, he_2024_035126_Phase}.
We set the lattice spacing $a$ to $1$.
In Fig.~\ref{fig:PD_CHIRAL}(a), at $D=0.5$, we find that the linear scaling of $1/\xi$ is perfectly satisfied in an extremely narrow region $J_2 - J^\text{Ising}_{2,\text{qc}} < 10^{-4}$ in the vicinity of the quasicritical point $J^\text{Ising}_{2,\text{qc}}$, yielding the exponent $\nu = 1$.
Next, we do an extrapolation with respect to the truncated bond dimension $m$ and find a transition point $J^\text{Ising}_{2,\text{c}}=0.408288(1)$ with a small error bar.
After defining a discrepancy of $\delta J^{\phantom{\dag}}_2 = J^{\phantom{\dag}}_2 - J^\text{Ising}_{2,\text{c}}$, we also do linear fitting of $\lvert \kappa \rvert$ in log-log scale in the region of $\ln \lvert \delta J_2 \rvert \lesssim -9$, which suggests $\lvert \kappa \rvert \propto \lvert \delta J_2 \rvert^\beta$ with $\beta=0.129$ close to $1/8$, as shown in Fig.~\ref{fig:PD_CHIRAL}(b).
In combination, we suggest that this transition is continuous and belongs to the Ising-type universality class.

In the inset of Fig.~\ref{fig:PD_CHIRAL}(b), we can also see the dimerization strength $\lvert \mathcal{D} \rvert$ does not disappear when $J_2 \ge J^\text{Ising}_{2,\text{c}}$, and the ground state enter the CPD phase, where the dimerization and vector chirality coexist.
Similarly, we can do the same procedure at the SHD-CSHD transition, which also belongs to the Ising type universality class, as shown in Fig.~\ref{fig:PD_CHIRAL}(c).
\begin{figure}[b]
\centering
\includegraphics[width=\columnwidth]{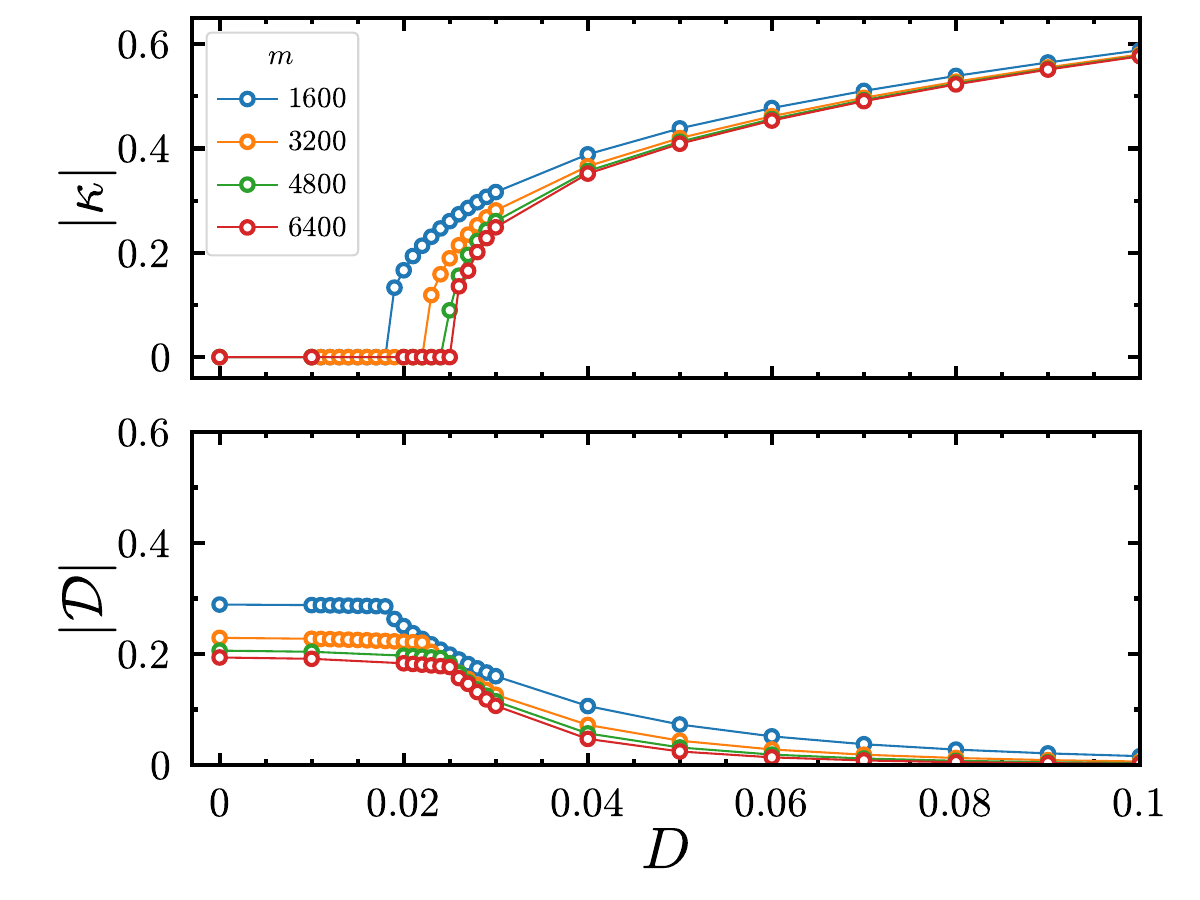}
\caption{(Color online)
(a) The vector chirality $\lvert \kappa \rvert$ and (b) the dimerization strength $\lvert \mathcal{D} \rvert$ as a function of $D$ for various $m$ at $J_2=4$.
\label{fig:CVBS}
}
\end{figure}
At $0.7 \lesssim D \le 1$, the effect of $D$ is counterbalanced by $J_2$, which induces a crossover from the CSHD phase to the CPD phase as $J_2$ increases, following the discussion in Sec.~\ref{subsec:zerodimerization}.
Furthermore, we make a trial to do extrapolation of the PD-CPD transition point $J^\text{Ising}_{2,c}$ to the limit of $D=0$ using parabolic function.
After that, we get a value of $J^\text{Ising}_{2,\text{c}} \simeq 0.55$ at $D=0$, close to the place of the PD-floating transition point determined by the early CFT analysis~\cite{chepiga_2020_174407_Floating}, although we do not fully understand the reasons behind this consistency.

The vector chiral order can be also generated from VBS.
As illustrated in Fig.~\ref{fig:CVBS}, as $D$ grows at $J_2 = 4$, we can see that the vector chirality strength $\lvert \kappa \rvert$ emerges abruptly with a finite discontinuity at a specific truncated bond dimension.
Meanwhile, the dimerization strength $\lvert \mathcal{D} \rvert$ does not vanish either, which leads to the last phase coexistence region, namely the CVBS phase.
It is recommended to adopt alternative initializations in order to avoid being trapped in a particular state during the iDMRG process.
The objective is to identify the state with the lowest energy per site.
With the growth of $m$ up to $6,400$, the jump persists.
Thus, it can be concluded that the transition may be of $1$st order in the limit of infinite $m$.

\subsection{Unstable floating phase}

\begin{figure}[b]
\centering
\includegraphics[width=\columnwidth]{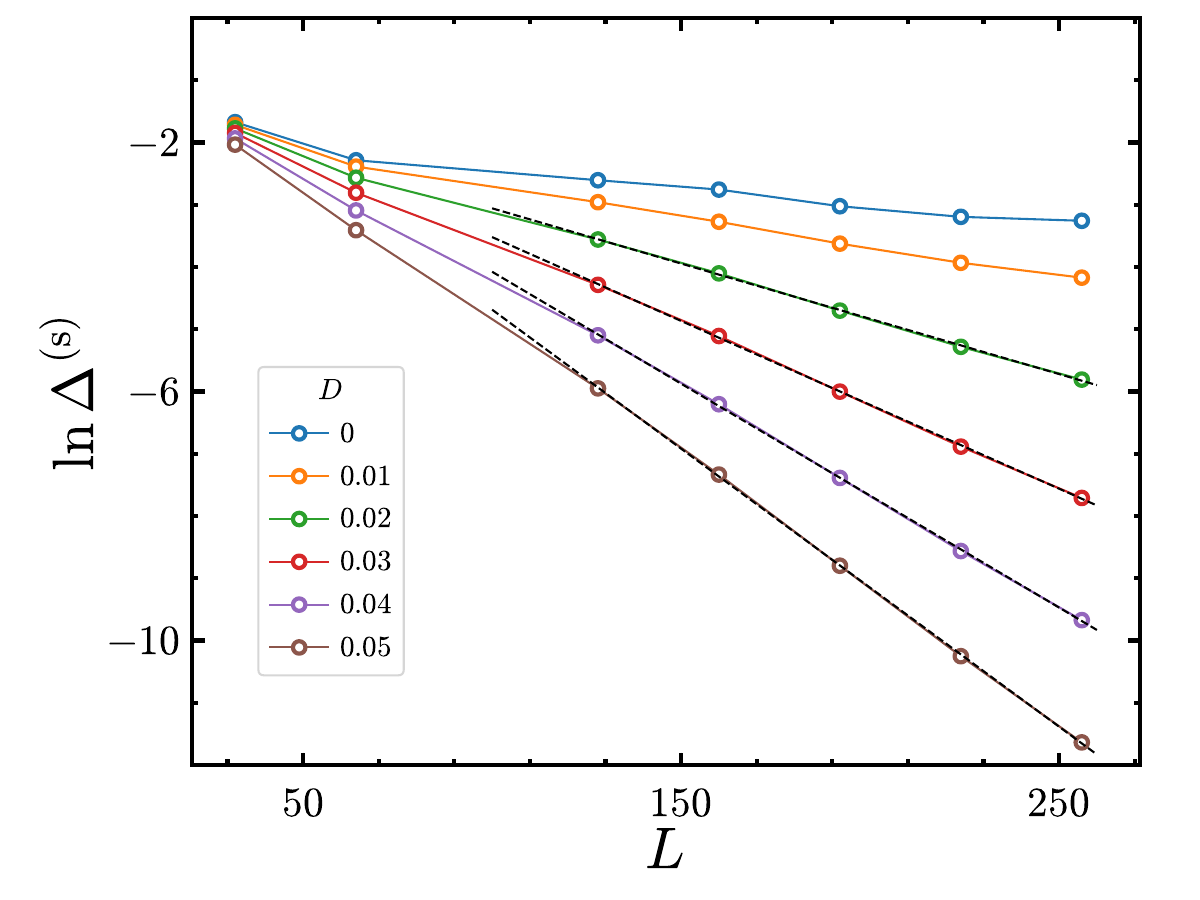}
\caption{(Color online)
The spin gap as a function of $L$ for various values of $D$ when $J_2=1$ is fixed.
The dashed lines indicate the linear functions of $L$ that best fit the gap values with $L=160$, $192$, $224$, and $256$ when $D \ge 0.02$.
$m=6,144$ and OBC are used in DMRG.
\label{fig:fl_chiral}
}
\end{figure}

After applying the uniaxial single ion anisotropy, we investigate the evolution of the floating phase by calculating the spin gap
\begin{equation}\label{eq:spingap}
\Delta^{(\text{s})} = E_\text{1st} - E_\text{g}\, ,
\end{equation}
where $E_\text{1st}$ represents the lowest-lying excited energy in the Hilbert space of $S^z_\text{t}=0$.
The floating phase is incommensurate and gapless, which is incompatible with the preinstalled condition for iDMRG~\cite{mcculloch_2008__Infinite,hu_2011_220402_Accurate}.
Thus we adopt the finite-chain algorithm in DMRG.
In Fig.~\ref{fig:fl_chiral}, the spin gap $\Delta^{(\text{s})}$ exhibits an exponential dependence on the chain length $L$, i.e., $\Delta^{(\text{s})} \propto \exp(- L/L_\text{h})$ when $D \gtrsim 0.02$.
This signal indicates that the energy discrepancy between two quasi-degenerate states diminishes as a consequence of the breaking of the space-inversion symmetry.
And $L_\text{h}$ represents the healing size.
Consequently, it can be posited that the floating phase is unstable and can undergo a transition to the vector chiral phase in the presence of a minimal amount of $D \lesssim 0.02$.

\subsection{CPD-CVBS, CPD-CSHD boundaries}\label{subsec:zerodimerization}

Analogous to the PD-SHD crossover mentioned in Sec.~\ref{subsec:PDvsSHD}, the edge and no-edge states of CPD, CSHD, and CVBS phases can be adiabatically connected separately.
Therefore, on the CPD-CSHD and CPD-CVBS boundary lines, the dimerization strength $\lvert \mathcal{D} \rvert$ disappear.
It follows that these three phases can be divided by two lines where $\lvert \mathcal{D} \rvert =0$.
In Fig.~\ref{fig:Phasediagram}, we use the non-edge states to determine the zero-dimerization lines when the truncated bond dimension $m=800$ is applied.
Two things are noticed: (\textbf{i}) the CPD-CSHD border is the extension of the PD-SHD crossover line for a slight growth of $J_2$, while the border accelerates the increasing along the $D$-axis when $J_2 \gtrsim 0.5$.
(\textbf{ii}) The extrapolation of the CPD-CVBS border to the limit of $D=0$ also hits the PD-floating transition point $J_2 \simeq 0.55$.

Moreover, in the CPD phase region, $\lvert \mathcal{D} \rvert$ is extremely weak, and it remains challenging to determine whether a pure chiral phase exists even using the most advanced techniques.
In the large-$D$ limit, the vector chiral order may persist due to the anisotropy parameter $\Delta_\text{eff} = 0.25$ being much smaller than one~\cite{nersesyan_1998_910_Incommensurate, hikihara_2001_174430_Groundstate} in the effective spin-$1/2$ $J_1$-$J_2$ $XXZ$ model as long as $J_2/J_1$ remains sufficiently large.
In that case, the CPD phase may end up before the ground state enters the CSHD phase region in the large-$D$ limit.

\vspace{1cm}

\section{Summary and conclusions}\label{sec:conclusion}
We have investigated the spin-$3/2$ $J_1$-$J_2$-$D$ model in the region with positive $D$ and $J_2$ using various methods.
In the groundstate phase diagram, a Luttinger liquid (LL) phase remains stable in the presence of $D>0$.
The frustration introduced by the $J_2$-term gives rise to the three distinct dimerized phases: the valence bond solid (VBS) phase, the partially dimerized (PD) phase, and the spin-$1/2$-like dimerized (SHD) phase, which can be distinguished by the spectra of reduced density matrices.
A crossover is found between the PD and SHD phases.
The Berezinskii-Kosterlitz-Thouless (BKT)-type transition lines between the LL, PD and SHD phases are determined using the level spectroscopy method.
As $J_2$ increases, regions of phase coexistence between dimerized and vector chiral orders emerges in a chiral PD (CPD) phase, a chiral SHD (CSHD) phase, and a chiral VBS (CVBS) phase.
The transitions between PD and CPD, and between SHD and CSHD are of Ising type, while the transition between VBS and CVBS is of $1$st order.
The floating phase rapidly vanishes when the vector chiral order appears upon introduction of $D$, illustrating its instability.
Finally, two zero-dimerization boundaries are identified, which partition three phase coexistence regions.

Several questions remain unanswered.
First, in the limit of $J_2=+\infty$, the competition between the VBS phase and the vector chiral phase may lead to a complicated scenario, the details of which are currently unclear.
Second, it is difficult to determine the exact boundaries of the floating phase and the VBS phase in the groundstate phase diagram.

\begin{acknowledgments}
We thank Ruizhen Huang for the helpful discussions. We acknowledge funding from the Ministry of Science and Technology of the People's Republic of China (Grant No.~2022YFA1402700) and the National Natural Science Foundation of China (Grants No.~U2230402).
S.~E. acknowledges funding from the Deutsche Forschungsgemeinschaft (DFG, German Research Foundation) Project No.~277625399-TRR 185 OSCAR (A5).
S.~J.~H. acknowledges funding from NSFC Grant No.~12174020.
The computations were performed on the Tianhe-2JK at the Beijing Computational Science Research Center (CSRC) and the Quantum Many-body {\rm I} cluster at the School of Physics and Astronomy, Shanghai Jiaotong University.
\end{acknowledgments}

\bibliography{ref}

\end{document}